\documentclass[conference]{IEEEtran}
\IEEEoverridecommandlockouts
\usepackage{cite}
\usepackage{amsmath,amssymb,amsfonts}
\usepackage{algorithmic}
\usepackage{graphicx}
\usepackage{textcomp}
\usepackage{xcolor}

\usepackage{hyperref}

\usepackage{kotex}
\usepackage{multirow}
\usepackage{multicol}
\usepackage{booktabs}

\newcommand{\ie}{\textit{i}.\textit{e}.}

\newcommand{\etal}{\textit{et}~\textit{al}.}

\def\BibTeX{{\rm B\kern-.05em{\sc i\kern-.025em b}\kern-.08em
    T\kern-.1667em\lower.7ex\hbox{E}\kern-.125emX}}
\begin{document}

\title{Inter-subject Contrastive Learning for Subject Adaptive EEG-based Visual Recognition

\thanks{
This work was supported by Institute for Information \& Communications Technology Planning \& Evaluation (IITP) grant funded by the Korea government (MSIT) (No. 2017-0-00451: Development of BCI based Brain and Cognitive Computing Technology for Recognizing Users Intentions using Deep Learning, No. 2020-0-01361: Artificial Intelligence Graduate School Program (YONSEI UNIVERSITY)).
}

}

\makeatletter
\def\footnoterule{\kern-3\p@
  \hrule \@width 2in \kern 2.6\p@} 
\makeatother

\author{\IEEEauthorblockN{Pilhyeon Lee}
\IEEEauthorblockA{\textit{Department of Computer Science} \\
\textit{Yonsei University} \\
Seoul, Republic of Korea \\
lph1114@yonsei.ac.kr}
\and
\IEEEauthorblockN{Sunhee Hwang}
\IEEEauthorblockA{\textit{AI Imaging Tech. Team} \\
\textit{LG Uplus}\\
Seoul, Republic of Korea \\
sunheehwang@lguplus.co.kr}
\and
\IEEEauthorblockN{Jewook Lee}
\IEEEauthorblockA{\textit{Department of Computer Science} \\
\textit{Yonsei University} \\
Seoul, Republic of Korea \\
hooraid@yonsei.ac.kr}
\and
\IEEEauthorblockN{Minjung Shin}
\IEEEauthorblockA{\textit{Graduate School of Artificial Intelligence} \\
\textit{Yonsei University} \\
Seoul, Republic of Korea \\
smj139052@yonsei.ac.kr}
\and
\IEEEauthorblockN{Seogkyu Jeon}
\IEEEauthorblockA{\textit{Department of Computer Science} \\
\textit{Yonsei University} \\
Seoul, Republic of Korea \\
jone9312@yonsei.ac.kr}
\and
\IEEEauthorblockN{Hyeran Byun\textsuperscript{*$\dagger$}}
\thanks{\textsuperscript{*}Corresponding author.}
\thanks{\textsuperscript{$\dagger$}Also with Graduate School of Artificial Intelligence and Graduate Program of Cognitive Science, Yonsei University.}
\IEEEauthorblockA{\textit{Department of Computer Science} \\
\textit{Yonsei University} \\
Seoul, Republic of Korea \\
hrbyun@yonsei.ac.kr}
}

\maketitle

\begin{abstract}
This paper tackles the problem of subject adaptive EEG-based visual recognition.
Its goal is to accurately predict the categories of visual stimuli based on EEG signals with only a handful of samples for the target subject during training.
The key challenge is how to appropriately transfer the knowledge obtained from abundant data of source subjects to the subject of interest.
To this end, we introduce a novel method that allows for learning subject-independent representation by increasing the similarity of features sharing the same class but coming from different subjects.
With the dedicated sampling principle, our model effectively captures the common knowledge shared across different subjects, thereby achieving promising performance for the target subject even under harsh problem settings with limited data.
Specifically, on the EEG-ImageNet40 benchmark, our model records the top-1 / top-3 test accuracy of 72.6\% / 91.6\% when using only five EEG samples per class for the target subject.
Our code is available at \href{https://github.com/DeepBCI/Deep-BCI/tree/master/1\_Intelligent\_BCI/Inter\_Subject\_Contrastive\_Learning\_for\_EEG}{https://github.com/DeepBCI/Deep-BCI}.
\end{abstract}


\begin{figure}[t]
    \centering
    \includegraphics[clip=true, width=0.80\columnwidth]{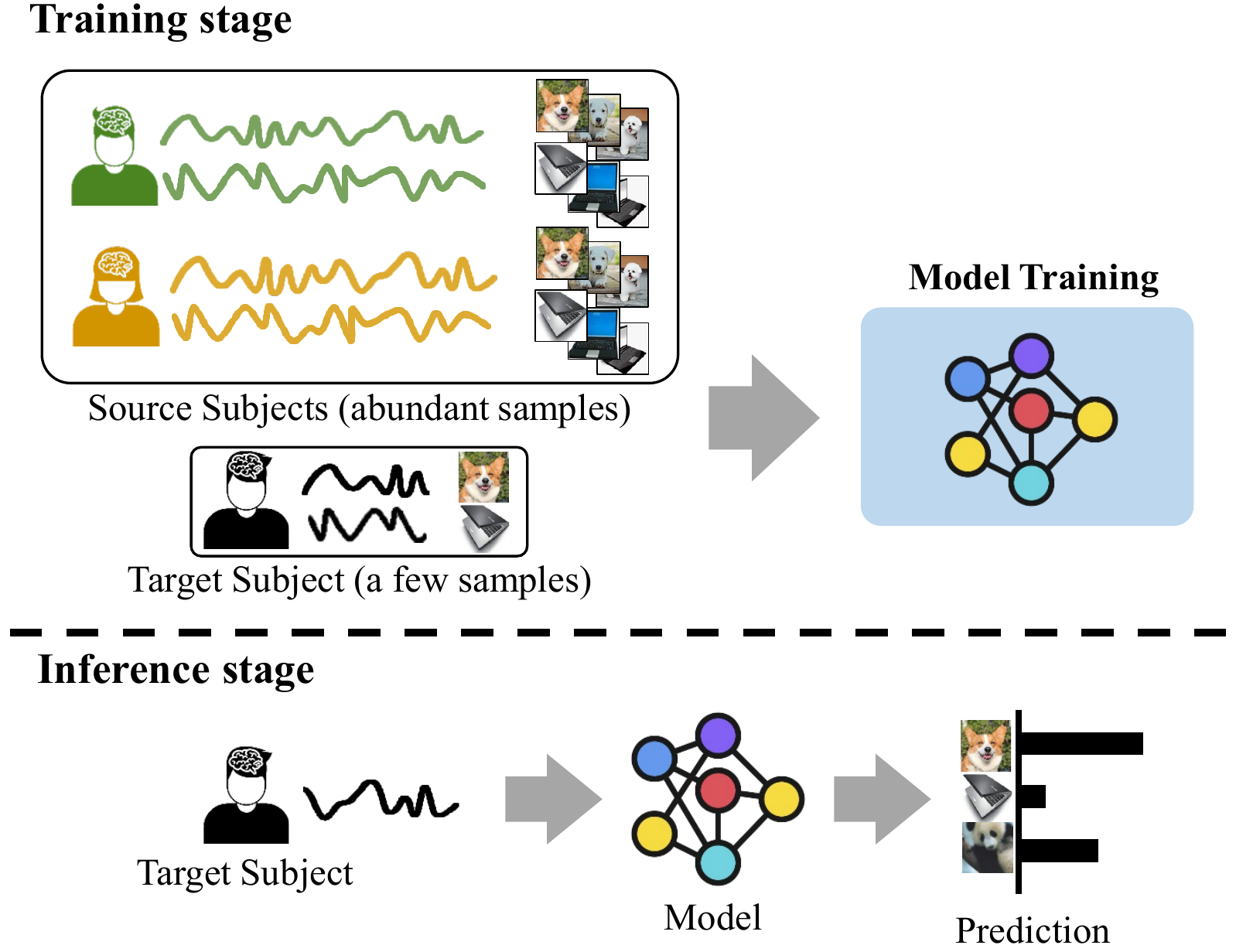}
    \caption{Graphical illustration of the problem setting. During training, we have abundant samples for source subjects but only a few samples for the target subject. At the inference stage, the model should correctly predict the visual category for EEG signals from the target subject.}
    \label{fig:intro_fig}
    \vspace{-3mm}
\end{figure}

\section{Introduction}
Reading the human mind is one of the most attracting problems for brain-computer interface (BCI) research.
Over the past decades, researchers have actively been studying how to decode the human mind from brain signals to facilitate communication or restore past memories~\cite{moment_restore,brain2image,thoughtviz,dream,decoding_bw}.
For example, BCI systems are developed to help people with disabilities to communicate with others by recognizing what they are thinking~\cite{brain_keyboard}.
In addition, researchers have paid much attention to interpreting visual information in the human mind and identifying the visual stimulus that a human has experienced from brain signals~\cite{fmri_memory}.
Recent studies have shown promising performance in recognizing the category of visual stimulus based on the brain signals~\cite{eeg_imgnet, lee2021subject-adaptive-EEG}.

There are several types of brain signals for developing BCI systems such as electroencephalography (EEG), magnetoencephalography (MEG), and functional magnetic resonance imaging (fMRI). 
Amongst, EEG is considered the most favorable one as it can be recorded non-invasively from the scalp of various locations by measuring the voltage fluctuations.
Owing to the convenience of the acquisition process, EEG-based BCI systems have already been largely adopted to the real-world applications of playing games~\cite{bci_game1,bci_game2} or brain training~\cite{training_brain}.
In the research fields, EEG-based classification tasks are also drawing attention, including emotion recognition~\cite{seed,seed_sbd}, drowsy detection~\cite{hwang2021bci,hwang2021icassp}, and decoding visual memory~\cite{eeg_imgnet, thoughtviz, decoding_bw}.
In this paper, we focus on the visual stimuli recognition task based on image-evoked EEG signals.

Although recent EEG-based visual recognition methods have shown great performance, most existing methods are developed in a subject-dependent way~\cite{mi1_sbd,mi2_sbd,seed_sbd,deap_sbd, mnist,obj_sbd}. 
Therefore, in order to utilize the existing methods for a new user, they necessitate the calibration process, \ie, re-training (or tuning) the model with an additionally collected dataset from the new user.
This process is labor-intensive and time-consuming, largely limiting the scalability of the EEG-based systems.
Most recently, a new practical problem setting is designed, named subject adaptive visual recognition~\cite{lee2021subject-adaptive-EEG}.
In this setting, a handful of samples are available for the target user, while abundant data for other subjects can be used during training.
At test time, the trained model is supposed to predict the visual categories based on EEG signals for the target user.
The problem setting is illustrated in Fig.~\ref{fig:intro_fig}.

The major challenge of the task is to appropriately learn the transferable knowledge across different subjects.
For the purpose, we propose a novel inter-subject contrastive loss that pulls the features coming from different subjects but sharing the same visual category with each other.
Moreover, we diagnose the inadequate factors in the conventional sampling method of contrastive learning and design a novel sampling strategy dedicated to subject-independent feature learning.
In the experiments on EEG-ImageNet40~\cite{eeg_imgnet}, we validate the effectiveness of the proposed method.
Our model consistently outperforms the existing methods including the vanilla approach with large margins under various settings.

We summarize our contributions as follows.
\begin{itemize}
    \item We propose the inter-subject contrastive loss that allows for learning the common knowledge across subjects for subject adaptive EEG-based visual recognition.
    \item We identify the inadequate properties in the sampling of contrastive learning and design a new sampling method customized for subject-independent feature learning.
    \item Through the quantitative analysis on EEG-ImageNet40, we verify the efficacy of our method. Specifically, it achieves the top-1 accuracy of 72.6\% in the 5-shot setting.
\end{itemize}

\section{Related Work}
\subsection{Brain activity underlying visual perception}
There have been a lot of attempts to interpret the brain activity~\cite{seed, deap_sbd,mi1_sbd,mi2_sbd,dd_2, dd_sbi, bci21_1,bci21_2}.
Specifically, a variety of studies are conducted for extracting visual information from brain waves recorded by diverse visual stimuli~\cite{fmri_memory,movie_recon,thoughtviz,brain2image,mnist}.
Some works tackle the problem of memory verification, whose goal is to identify whether the visual event is occurred or not in one's experience based on the brain signals~\cite{fmri_memory,face_memory}.
Meanwhile, other works try to reconstruct the visual stimuli from brain signals such as object images~\cite{thoughtviz,brain2image} and movie scenes~\cite{movie_recon}.
Furthermore, a bunch of methods have been proposed to recognize the visual category of brain signals recorded while seeing digits~\cite{mnist} or object images~\cite{eeg_imgnet}.
Most recently, a realistic problem setting for visual recognition is proposed, namely subject adaptive EEG-based visual recognition~\cite{lee2021subject-adaptive-EEG}.
In this work, we tackle the practical subject adaptive visual recognition task based on EEG signals induced by the visual object images.

\subsection{Subject-independent EEG-based classification}
To achieve the generalization ability of EEG-based classification model, researchers have focused on developing the subject-independent models.
To this end, some works attempt to extract subject-independent features from EEG signals~\cite{feature_sbi,feature2_sbi}.
On the other hand, several approaches adopt the adversarial training strategy~\cite{em_sbi,sd_sbi,dd_sbi} to train the model to classify the labels correctly, but not to distinguish the subject identities.
Moreover, there are several methods~\cite{lee2021subject-adaptive-EEG, li2019domain,da_aaai} that adopt existing domain adaptation methods to facilitate adapting the model to a new target user. 
In contrast to the works above, we propose a novel contrastive learning strategy for subject-independent feature learning for subject adaptive EEG-based visual recognition.

\subsection{Contrastive learning}
Recently, the field of unsupervised learning has rapidly grown with the powerful contrastive learning\cite{oord2018representation,simclr,moco,multiview,pretextinvariant}.
It enables deep models to learn informative representation by attracting positive pairs from an anchor and pushing negative pairs away from it~\cite{oord2018representation}.
Various techniques for contrastive learning such as momentum encoder\cite{moco} and data augmentation strategies\cite{simclr} have enhanced the representation ability and and the training efficiency. Furthermore, Grill~\etal\cite{grill2020bootstrap} overcome the limitation of contrastive learning which requires a huge batch size for promising performance, by proposing a negative cosine similarity loss with the stop-gradient operation.
Meanwhile, Khosla~\etal\cite{Supcon} extend the contrastive loss to the supervised setting. By involving all samples of the same class with an anchor in the positive bag and pulling them from an anchor, they learn a more discriminative representation for the classes.
In this paper, we propose to utilize contrastive learning for learning subject-independent representation and design a new positive/negative sampling strategy.

\section{Method}
\label{sec:method}
\subsection{Preliminaries}
Our goal is to effectively train a model on a training set that consists of only a few samples for the target subject and abundant data for the source subjects.
Let $\mathcal{D}^{train}=\mathcal{D}^{src} \cup \mathcal{D}^{trg}$ be the whole training set, where $\mathcal{D}^{src}$ and $\mathcal{D}^{trg}$ indicate the datasets from the source and target subjects, respectively.
Here, we denote the source dataset by the union of multiple datasets from different subjects, \ie, $\mathcal{D}^{src}=\bigcup_{j=1}^{S}\mathcal{D}^{src}_{j}$, where $\mathcal{D}^{src}_{j}$ indicates the set of the EEG signals collected from the $j$-th subject, $S$ is the total number of source subjects, and $\bigcup$ is the set union operator.
According to the problem formulation, the dataset of each source subject contains a large number of signals, \ie, $|\mathcal{D}^{src}_{j}| \gg 0$.
On the other hand, we have only a few EEG samples of the target subject for training; that is, $|\mathcal{D}^{trg}| \ll |\mathcal{D}^{src}_{j}|$ for all $j \in [1, S]$.
Each training example is denoted by a triplet, $\{x_i, s_i, y_i\} \in \mathcal{D}^{train}$, where $x_i \in \mathbb{R}^{D \times T}$ indicates the $i$-th sample with its duration of $T$ and the channel dimension of $D$, while $s_i \in \mathbb{R}^{S+1}$ and $y_i \in \mathbb{R}^{K}$ respectively are the subject identity and the class label of the $i$-th sample.
After training the model on $\mathcal{D}^{train}$, we evaluate it on an unseen dataset $\mathcal{D}^{test}$ that is \textit{exclusively} collected from the target subject.

In practice, instead of utilizing randomly sampled batches from the training set, we construct balanced batches of samples for stable model training.
Specifically, we randomly sample $N$ samples for each subject including sources and the target, which then constitute a single batch $\mathcal{B}$ where $|\mathcal{B}|=(S+1)N$.
Note that the target dataset may have insufficient samples, \ie, $|\mathcal{D}^{trg}|<N$.
In this case, we duplicate them to simulate a total of $N$ samples, which is also known as oversampling.

\subsection{Architecture}
\label{sec:architecture}
Given an input signal $x_i \in \mathbb{R}^{D \times T}$, our model is supposed to predict its class probabilities $p(\mathbf{y}|x) \in \mathbb{R}^{K}$, where $D$ and $T$ represent the channel dimension and the duration of the input respectively, while $K$ is the number of visual classes.
Following the previous work~\cite{lee2021subject-adaptive-EEG}, we build a model that consists of a sequence encoder $f$ and an embedding layer $g$ followed by a classifier $h$.
Specifically, the sequence encoder $f(\cdot)$ is a single layer gated recurrent unit (GRU) for temporal modeling, while the embedding layer $g(\cdot)$ and the classifier $h(\cdot)$ are a fully-connected (FC) layer with activation functions.
Taking an EEG signal as input, the sequence encoder generates the encoded feature representation $z_i=f(x_i) \in \mathbb{R}^{D_{enc}}$, where $D_{enc}$ denotes the dimension of the encoded feature.
We keep the output feature at the last timestamp and discard the other intermediate ones.
The encoded feature is then fed into the embedding layer, resulting in the embedded feature $w_i=g(z_i) \in \mathbb{R}^{D_{emb}}$, where $D_{emb}$ is the dimension of the embedded feature.
Lastly, the classifier predicts the class probabilities given the embedded features, \ie, $p(\mathbf{y}_i|x_i;\theta)=h(w_i) \in \mathbb{R}^{K}$.
Here, $\theta$ denotes all the trainable parameters in our model.

\begin{figure}[t]
    \centering
    \includegraphics[clip=true, width=1\columnwidth]{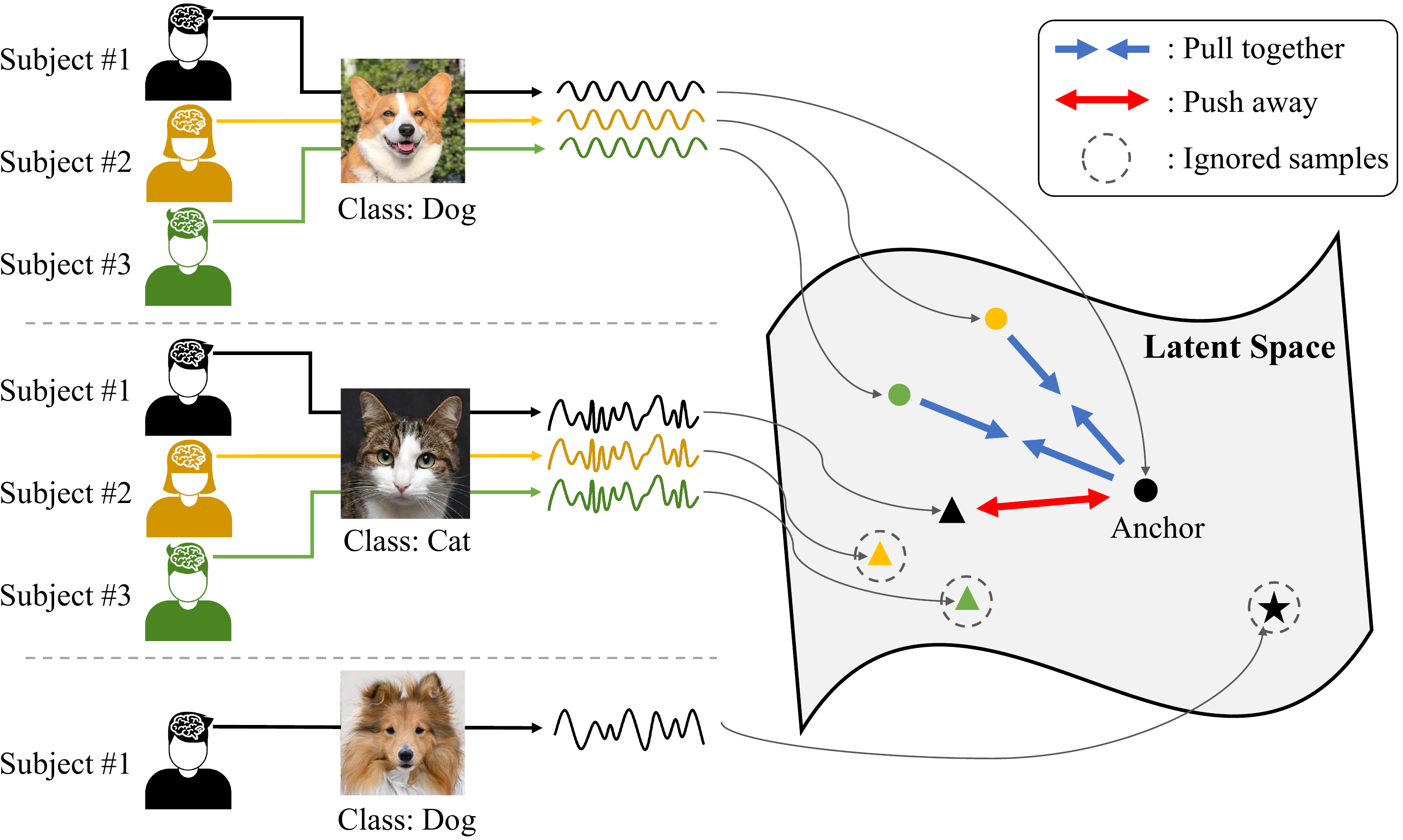}
    \caption{Illustration of our inter-subject contrastive learning.}
    \label{fig:contrast_fig}
    \vspace{-1mm}
\end{figure}

\subsection{Training objectives}
We utilize two loss functions for model training, namely the classification loss and the inter-subject contrastive loss.
The former is for the classification task, while the latter is what we propose for subject-independent feature learning.

\subsubsection{Classification loss}
We adopt the conventional cross-entropy loss for the classification task, which is formulated as follows.
\begin{equation}
    \mathcal{L}_{\text{cls}} = -  \frac{1}{|\mathcal{B}|}\sum_{\forall i \in \mathcal{B}} y_i \log p(y_i|x_i;\theta),
    \label{equ:cross_entropy}
\end{equation}
where $\mathcal{B}$ indicates the training batch at each training step, which contains the indices of the samples in the batch.

\subsubsection{Inter-subject contrastive loss}
To allow the model to learn invariant feature representation to different subjects, we propose a novel inter-subject contrastive loss.
In detail, the loss is designed to encourage the similarity between features sharing the same class but coming from different subjects.
Formally, the loss is computed by:
\begin{equation}
    \mathcal{L}_{\text{contrast}} = - \frac{1}{|\mathcal{B}|}\sum_{\forall i \in \mathcal{B}}
    \log
    \frac{\sum_{\forall j \in \mathcal{P}(i)}
    \exp(z_i \cdot z_j / \tau)}
    {\sum_{\forall k \in \mathcal{A}(i)}
    \exp(z_i \cdot z_k / \tau)},
    \label{equ:contrastive}
\end{equation}
where $\mathcal{B}$ indicates the training batch, while $\mathcal{A}(i)$ and $\mathcal{P}(i)$ respectively denote the anchor set and the positive set with regard to the sample $x_i$ (\ie, anchor).
$\tau$ is a temperature parameter that controls the sharpness of the similarity distribution.
Here an important question is, how should we define the anchor and the positive sets?

In the conventional supervised contrastive loss~\cite{Supcon}, the anchor set is defined by the other samples in the batch, \ie, $\mathcal{A}(i)=\mathcal{B} \setminus \{i\}$ and the positive set is composed of the samples sharing the same class, \ie, $\mathcal{P}(i)=\{j \in \mathcal{A}: y_i = y_j\}$.
Accordingly, the negative samples are defined by those not contained in the positive set, \ie, $\mathcal{N}(i)=\mathcal{A}(i) \setminus \mathcal{P}(i)$.
After defining the sets, the contrastive loss pulls the anchor and the positive samples together while pushing away the anchor from the negative samples.
Despite the strong performance in classification, we observe that this sampling strategy is inadequate for subject-independent feature learning in two aspects:
\begin{itemize}
    \item Due to the low intra-subject variation, the loss value is prone to be dominated by the positive samples obtained from the same subject, \ie, $x_j \in \mathcal{P}(i)$ such that $y_i=y_j$ and $s_i=s_j$.
    \item Pushing away the negative samples of different subjects from the anchor, \ie, $x_j \in \mathcal{N}(i)$ such that $y_i \neq y_j$ and $s_i \neq s_j$, hinders the subject-independent feature learning by increasing the feature discrepancy between different subjects.
\end{itemize}
Based on these observations, we design a dedicated sampling strategy for subject-adaptive feature learning.
In specific, we ignore such samples when building the anchor set.
The resulting anchor set is defined by $\mathcal{A}'(i)=\mathcal{A}_1^{sub}(i) \cup \mathcal{A}_2^{sub}(i)$ with the subsets $\mathcal{A}_1^{sub}(i)=\{ j \in \mathcal{B} : y_i = y_j, s_i \neq s_j\}$ and $\mathcal{A}_2^{sub}(i)=\{ j \in \mathcal{B} : y_i \neq y_j, s_i = s_j\}$.
Accordingly, the positive and the negative samples are decided based on whether they have the same classes with the anchor, \ie, $\mathcal{P}'(i)=\mathcal{A}_1^{sub}(i)$ and $\mathcal{N}'(i)=\mathcal{A}_2^{sub}(i)$.
We illustrate the inter-subject contrastive learning process equipped with our sampling strategy in Fig.~\ref{fig:contrast_fig}.

Our model is jointly trained with the two loss functions.
The overall loss is computed by:
\begin{equation}
    \mathcal{L}_{\text{total}} = \mathcal{L}_{\text{cls}} + \lambda\mathcal{L}_{\text{contrast}},
\end{equation}
where $\lambda$ is a weighting coefficient for balancing the losses, which we empirically set to 1.

\section{Experiments}
\label{sec:experiments}
In this section, we first describe the specifications of the EEG-signal dataset~\cite{eeg_imgnet}. Then we report the implementation details of the proposed method. Afterwards, we quantitatively estimate the performance with both top-1 and top-3 accuracy. Furthermore, we conduct extensive analysis on our method to demonstrate the effectiveness.

\subsection{Dataset}
For experiments, we utilize a public large-scale dataset, EEG-ImageNet40~\cite{eeg_imgnet}. It consists of 128-channel EEG signals with a duration of 440 ms, which are collected from six different subjects: one female and five male. To remove noisy signals and capture the frequency bands that are related to visual perception (\ie, Beta and Gamma), a notch filter (49-51Hz) and a band-pass filter (14-72 Hz) are used for filtering EEG signals. The dataset is composed of 40 object categories from ImageNet~\cite{ImageNet}, each class consisting of roughly 50 images. For evaluating, we follow the ratio of 4:1:1 provided by the official dataset contributor for training:validation:test. We follow the official 6 splits for experiments and report the mean and the standard deviation of results from all splits.


\begin{table}[t]
\caption{
Quantitative comparison by changing the target subject in the 1-shot setting.
}
\centering
\resizebox{1.0\columnwidth}{!}{
\begin{tabular}{c|ccc|ccc}
\toprule
\multicolumn{7}{c}{Validation set} \\ \midrule
\multirow{2}{*}{Subject}  & \multicolumn{3}{c|}{top-1 accuracy (\%)} & \multicolumn{3}{c}{top-3 accuracy (\%)} \\ 
       & Vanilla     & MMD~\cite{lee2021subject-adaptive-EEG}  & Ours   & Vanilla     & MMD~\cite{lee2021subject-adaptive-EEG}  & Ours  \\ \midrule
\#0 & $29.3_{\pm1.9}$   & $35.7_{\pm1.9}$    & $\textbf{39.2}_{\pm1.8}$   & $51.6_{\pm3.0}$   & $58.1_{\pm2.9}$   & $\textbf{66.9}_{\pm3.2}$   \\
\#1 & $21.8_{\pm2.3}$   & $29.0_{\pm3.6}$    & $\textbf{33.0}_{\pm2.6}$   & $41.0_{\pm5.1}$   & $49.5_{\pm3.5}$   & $\textbf{54.1}_{\pm2.3}$   \\
\#2 & $25.3_{\pm0.9}$   & $30.8_{\pm2.2}$    & $\textbf{36.9}_{\pm2.2}$   & $44.4_{\pm2.1}$   & $53.1_{\pm2.6}$   & $\textbf{62.5}_{\pm2.9}$   \\
\#3 & $28.8_{\pm2.2}$   & $31.9_{\pm3.9}$    & $\textbf{43.6}_{\pm4.7}$& $47.8_{\pm4.1}$   & $52.6_{\pm3.7}$   & $\textbf{61.2}_{\pm4.2}$   \\
\#4 & $25.9_{\pm1.9}$   & $36.2_{\pm3.3}$    & $\textbf{40.0}_{\pm3.3}$   & $44.4_{\pm2.7}$   & $61.0_{\pm4.7}$   & $\textbf{65.2}_{\pm2.2}$   \\
\#5 & $20.7_{\pm2.9}$   & $25.8_{\pm1.7}$    & $\textbf{34.8}_{\pm1.9}$   & $40.1_{\pm3.9}$   & $47.5_{\pm3.4}$   & $\textbf{59.9}_{\pm2.4}$  \\ \midrule

\multicolumn{7}{c}{Test set} \\ \midrule
\multirow{2}{*}{Subject}  & \multicolumn{3}{c|}{top-1 accuracy (\%)} & \multicolumn{3}{c}{top-3 accuracy (\%)} \\ 
       & Vanilla     & MMD~\cite{lee2021subject-adaptive-EEG}   & Ours   & Vanilla     & MMD~\cite{lee2021subject-adaptive-EEG}   & Ours  \\ \midrule
\#0 & $24.3_{\pm0.9}$   & $29.6_{\pm4.9}$    & $\textbf{32.8}_{\pm4.5}$   & $48.3_{\pm2.3}$   & $56.8_{\pm4.1}$   & $\textbf{64.4}_{\pm6.2}$   \\
\#1 & $18.1_{\pm2.7}$   & $25.4_{\pm2.4}$    & $\textbf{29.2}_{\pm3.1}$   & $39.0_{\pm1.9}$   & $49.0_{\pm2.4}$   & $\textbf{55.1}_{\pm1.6}$   \\
\#2 & $23.9_{\pm3.0}$   & $29.2_{\pm3.7}$    & $\textbf{35.6}_{\pm3.8}$   & $44.3_{\pm4.3}$   & $54.5_{\pm3.1}$   & $\textbf{62.7}_{\pm5.0}$   \\
\#3 & $27.4_{\pm3.2}$   & $32.1_{\pm4.3}$    & $\textbf{40.4}_{\pm4.2}$   & $47.9_{\pm4.2}$   & $53.6_{\pm4.0}$   & $\textbf{59.5}_{\pm5.3}$   \\
\#4 & $22.7_{\pm1.2}$   & $35.3_{\pm3.6}$    & $\textbf{37.5}_{\pm3.7}$   & $44.8_{\pm3.5}$   & $60.7_{\pm4.9}$   & $\textbf{65.8}_{\pm3.1}$   \\
\#5 & $18.9_{\pm2.9}$   & $21.4_{\pm2.6}$    & $\textbf{30.2}_{\pm2.9}$   & $38.4_{\pm4.1}$   & $45.0_{\pm4.1}$  & $\textbf{58.7}_{\pm2.1}$   \\
    \bottomrule
\end{tabular}
}
\label{table:quant_subject}
\end{table}

\subsection{Implementation details}
Based on the analysis of the original paper~\cite{eeg_imgnet}, we use the signals with the interval of 320-480~\textit{ms}. The signals have 128 channels (\ie, $D=128$) and its temporal dimension becomes 160 after temporal clipping, \ie, $T=160$. The dimensions of features, \ie, $D_{enc}$ and $D_{emb}$, are set to 128 equally.
The temperature parameter $\tau$ of our inter-subject contrastive loss is set to $0.05$. During training, the number of samples for each subject in a batch is set to 200 (\ie, $N=200$), leading to the total batch size of 1,200. For parameter optimization, we adopt the Adam optimizer~\cite{kingma2014adam}, and the learning rate is set to $0.001$. We train our model in an end-to-end manner from scratch.

\subsection{Quantitative results}
To demonstrate the effectiveness of our method, we quantitatively compare our method with the vanilla baseline and the previous method~\cite{lee2021subject-adaptive-EEG}.
The vanilla baseline is an ablated version of our method, which only has the classification loss as its training objective.
Following the previous work~\cite{lee2021subject-adaptive-EEG}, the comparisons are made in k-shot settings. In the $k$-shot setting, only $k$ samples of target subjects for each class are available with abundant source data during training. In order to evaluate the methods under various settings, we vary $k$ from 1 to 5 and compare their performance. Note that the 1-shot setting is the most extreme case of $k$-shot scenario where only a single target sample per class is accessible. We adopt top-1 and top-3 accuracy as the evaluation metrics.

\begin{table}[t]
\caption{
Quantitative comparison by changing the number of target samples per class used for training.
}
\centering
\resizebox{0.93\columnwidth}{!}{
\begin{tabular}{c|ccc|ccc}
\toprule
\multicolumn{7}{c}{Validation set} \\ \midrule
\multirow{2}{*}{$k$}  & \multicolumn{3}{c|}{top-1 accuracy (\%)} & \multicolumn{3}{c}{top-3 accuracy (\%)} \\ 
       & Vanilla     & MMD~\cite{lee2021subject-adaptive-EEG}      & Ours   & Vanilla     & MMD~\cite{lee2021subject-adaptive-EEG}      & Ours  \\ \midrule
1 & $25.3_{\pm1.0}$   & $31.7_{\pm1.5}$    & $\textbf{37.9}_{\pm4.6}$   & $44.9_{\pm1.3}$   & $53.6_{\pm1.9}$   & $\textbf{61.6}_{\pm5.1}$   \\
2 & $41.7_{\pm1.9}$   & $46.3_{\pm1.8}$    & $\textbf{54.3}_{\pm5.8}$   & $65.2_{\pm2.0}$   & $70.2_{\pm1.6}$   & $\textbf{77.8}_{\pm3.8}$   \\
3 & $54.4_{\pm1.0}$   & $58.9_{\pm0.7}$    & $\textbf{64.7}_{\pm5.2}$   & $77.6_{\pm0.7}$   & $80.8_{\pm1.2}$   & $\textbf{86.1}_{\pm3.4}$   \\
4 & $64.6_{\pm1.5}$   & $67.5_{\pm1.2}$    & $\textbf{71.6}_{\pm5.4}$   & $85.1_{\pm1.1}$   & $86.8_{\pm1.2}$   & $\textbf{89.6}_{\pm3.2}$   \\
5 & $72.0_{\pm1.3}$   & $73.5_{\pm1.1}$    & $\textbf{76.1}_{\pm4.8}$   & $89.6_{\pm0.9}$   & $90.0_{\pm1.0}$  & $\textbf{92.1}_{\pm1.7}$   \\ \midrule
\multicolumn{7}{c}{Test set} \\ \midrule
\multirow{2}{*}{$k$}  & \multicolumn{3}{c|}{top-1 accuracy (\%)} & \multicolumn{3}{c}{top-3 accuracy (\%)} \\  
       & Vanilla     & MMD~\cite{lee2021subject-adaptive-EEG}      & Ours   & Vanilla     & MMD~\cite{lee2021subject-adaptive-EEG}      & Ours  \\ \midrule
1 & $22.5_{\pm0.8}$   & $28.8_{\pm1.2}$    & $\textbf{34.3}_{\pm5.5}$   & $43.8_{\pm1.6}$   & $53.3_{\pm1.9}$   & $\textbf{61.0}_{\pm5.6}$   \\
2 & $39.9_{\pm2.0}$   & $43.8_{\pm1.4}$    & $\textbf{51.2}_{\pm6.1}$   & $65.1_{\pm2.1}$   & $69.5_{\pm1.4}$   & $\textbf{77.5}_{\pm3.9}$   \\
3 & $52.6_{\pm1.7}$   & $56.4_{\pm1.7}$    & $\textbf{62.2}_{\pm6.3}$   & $77.0_{\pm1.5}$   & $80.4_{\pm1.1}$   & $\textbf{85.6}_{\pm3.6}$   \\
4 & $62.4_{\pm1.7}$   & $64.7_{\pm1.6}$    & $\textbf{68.6}_{\pm5.9}$   & $84.3_{\pm0.9}$   & $85.9_{\pm1.1}$   & $\textbf{89.1}_{\pm3.5}$   \\
5 & $69.5_{\pm1.1}$   & $70.1_{\pm1.0}$    & $\textbf{72.6}_{\pm5.1}$   & $89.0_{\pm0.5}$   & $89.2_{\pm0.5}$  & $\textbf{91.6}_{\pm2.5}$   \\
    \bottomrule
\end{tabular}

}
\label{table:quant_k}
\end{table}

\subsubsection{Comparison in the 1-shot setting}
In Table~\ref{table:quant_subject}, we conduct a comparative study on different target subjects in 1-shot setting.
Regardless of the target subject choice, our method records the best accuracy.
The vanilla setting shows poor performance for all target subjects, which indicates that the large inter-subject variation prevents the abundant source data from improving the performance for the target subject.
As it is trained only with the cross-entropy loss, the learned representation is dependent to source subjects and thus is not helpful for other subjects.
In addition, our method outperforms MMD~\cite{lee2021subject-adaptive-EEG} that is adoption of a domain adaptation technique, demonstrating the effectiveness of the proposed inter-subject contrastive loss in learning subject-independent representations. Especially, the top-1 accuracy on validation set of subject \#$5$ is largely improved by 9.0\% from MMD~\cite{lee2021subject-adaptive-EEG}.

\subsubsection{Comparison with varying k}
In Table~\ref{table:quant_k}, we provide comparisons with varying numbers of target samples per class for training. Notably, our method surpasses the performance of both vanilla and MMD model regardless of $k$. To be more specific, the performance gap between the competitors and our method becomes more significant as k gets smaller (more harsh settings). The overall results verify that the proposed inter-subject contrastive loss successfully reduces the inter-subject discrepancy, which is even more effective when less target samples are available.
On the other hand, the performance gain become smaller as $k$ grows.
We however note that collecting a large volume of target subject signals is costly and impractical in many real-world scenarios.

\subsection{Analysis on the sampling method}
In Sec.~\ref{sec:method}, we argued that the conventional sampling method of contrastive learning conflicts with our goal, \ie, subject-independent feature learning.
Accordingly, we proposed a dedicated sampling strategy that better suits our goal.
To verify the claim, we compare our method with the conventional sampling on the validation set of EEG-ImageNet40 in the 1-shot setting.
The results are shown in Table~\ref{table:analysis_contrastive}.
It is obviously shown that our method performs significantly better than the conventional sampling.
This result indicates that the existing contrastive sampling method could be sub-optimal for learning subject-independent representation and a carefully designed sampling strategy like ours is desired.


\begin{table}[t]
\caption{
Comparison with the conventional sampling method of contrastive learning
}
\centering
\resizebox{0.9\columnwidth}{!}{
\begin{tabular}{c|cc|cc}
\toprule
\multirow{2}{*}{Subject}  & \multicolumn{2}{c|}{top-1 accuracy (\%)} & \multicolumn{2}{c}{top-3 accuracy (\%)} \\ 
       & Conventional~\cite{Supcon}  & Ours   & Conventional~\cite{Supcon}  & Ours  \\ \midrule
\#0 & $30.1_{\pm1.0}$    & $\textbf{39.2}_{\pm1.8}$   & $51.3_{\pm1.7}$   & $\textbf{66.9}_{\pm3.2}$   \\
\#1 & $29.2_{\pm3.0}$    & $\textbf{33.0}_{\pm2.6}$   & $47.8_{\pm3.7}$   & $\textbf{54.1}_{\pm2.3}$   \\
\#2 & $29.2_{\pm1.3}$    & $\textbf{36.9}_{\pm2.2}$   & $49.3_{\pm2.7}$   & $\textbf{62.5}_{\pm2.9}$   \\
\#3 & $38.1_{\pm3.9}$    & $\textbf{43.6}_{\pm4.7}$   & $56.7_{\pm4.2}$   & $\textbf{61.2}_{\pm4.2}$   \\
\#4 & $29.4_{\pm1.4}$    & $\textbf{40.0}_{\pm3.3}$   & $46.0_{\pm1.8}$   & $\textbf{65.2}_{\pm2.2}$   \\
\#5 & $22.5_{\pm2.1}$    & $\textbf{34.8}_{\pm1.9}$   & $51.7_{\pm2.3}$   & $\textbf{59.9}_{\pm2.4}$  \\ 
    \bottomrule
\end{tabular}
}
\label{table:analysis_contrastive}
\end{table}


\section{Conclusion}
In this paper, we have presented a new approach for subject adaptive EEG-based visual recognition.
In order to learn the shared knowledge across different subjects, we proposed the inter-subject contrastive loss.
In addition, we also designed a sampling strategy that well suits the subject-independent representation learning.
In the experiments under various environments, the effectiveness of our method is clearly validated showing promising results in the data scarcity settings.
As future work, we would like to explore more practical problem settings such as subject generalization~\cite{ghifary2015domain,jeon2021stylizationDG}.

\bibliographystyle{IEEEtran}
\bibliography{ref}

\end{document}